\documentclass[]{article}
\usepackage{spconf,amsmath,epsfig, amsfonts, amssymb}
\def\X{{\mathbf X}}

\newcommand{\marpas}{Mar\v{c}enko-Pastur}

\newcommand{\GNL}{\ensuremath{{\bf \mathcal G}}}

\newcommand{\Lmz}{\ensuremath{L_{mz}}}

\newcommand{\Knn}{\ensuremath{\Cov_{X}}}

\newcommand{\herm}{^{H}}

\newcommand{\WishartUL}{\ensuremath{(1 + \sqrt{c})^2}}
\newcommand{\Threshold}{\ensuremath{(1 + \sqrt{c})}}
\newcommand{\edf}{EDF}
\newcommand{\cin}{cylindrically isotropic noise}
\newcommand{\ECM}{ECM}
\newcommand{\fMP}{\ensuremath{f_{MP(c)}}} 
\newcommand{\fhat}{\ensuremath{\hat{f}(x)}}
\newcommand{\Wishart}{\ensuremath{{\bf W}(c)}}
\newcommand{\Cov}{\ensuremath{\boldsymbol{\Sigma}}}

\newcommand{\eval}{\ensuremath{\gamma}}
\newcommand{\seval}{\ensuremath{g}}

\newcommand{\sampCov}{\ensuremath{{\bf S}_X}}

\newcommand{\datamtx}{\ensuremath{{\bf X}}}

\title{APPROXIMATE EIGENVALUE DISTRIBUTION OF A CYLINDRICALLY ISOTROPIC NOISE SAMPLE COVARIANCE MATRIX}
\twoauthors
  {Saurav R. Tuladhar, John R. Buck\thanks{SRT and JRB were funded by ONR Awards	N00014-09-1-167 and N00014-12-1-0047. KEW was funded by ONR Awards N00014-09-1-0114 and N00014-12-1-0048.}}
	{University of Massachusetts\\
	ECE Dept.\\
	N. Dartmouth, MA 02740}
  {Kathleen E. Wage} 
	{George Mason University\\
	ECE Dept.\\
	Fairfax, VA 22030}
\begin{document}
%
\maketitle
\begin{abstract}
The statistical behavior of the eigenvalues of the sample covariance matrix (SCM) plays a key role in determining the performance of adaptive beamformers (ABF) in presence of noise. This paper presents a method to compute the approximate eigenvalue density function (EDF) for the SCM of a \cin{} field when only a finite number of shapshots are available. The EDF of the ensemble covariance matrix (ECM) is modeled as an atomic density with many fewer atoms than the SCM size. The model results in substantial computational savings over more direct methods of computing the EDF. The approximate EDF obtained from this method agrees closely with histograms of eigenvalues obtained from simulation.

\end{abstract}
\section{Introduction}
\label{sec:intro}
In array processing, adaptive beamformers (ABF) rely on the knowledge
of the spatial covariance matrix of the data \cite{vtree08array}.  In
most applications the ensemble covariance matrix (ECM) is not known a
priori, thus it must be estimated from measurements. A common
technique for estimating the ECM is to compute the sample covariance
matrix (SCM).

A spatially white background noise is a common assumption in analyzing the 
performance of ABFs in presence of noise. In practice however, a spatially 
correlated noise field may exist in the environment. In a shallow underwater 
acoustic channel, the correlated noise is generally modeled as cylindrically  
isotropic field \cite{cox1973spatial}. The noise model developed in \cite{kuperman80spatial}
simplifies to cylindrically isotropic noise for a horizontal linear array
at a constant depth.


Assuming a uniform linear array placed on the plane of symmetry of 
the noise field, the entries of the ECM (\Knn{}) for the \cin{} field are 
given by 
\begin{equation}
[\Knn]_{pq} = J_0(2\pi\zeta|p - q|)  \label{eqn:iso_noise_cov},
\end{equation}
where $J_0()$ is the zeroth order Bessel function of the first kind and $\zeta$ is the ratio of the sensor spacing to wavelength. The statistical behavior of the eigenvalues and eigenvectors of the SCM in the presence of noise plays a crucial role in the performance of ABFs. Thus, understanding the distribution of the eigenvalues of the noise SCM is important for ABFs.

Traditionally the replacement of the ECM by the SCM in ABFs was justified by the asymptotic convergence of the SCM to the ECM. However, in practice the SCM has to be estimated from a finite number of snapshots. The number of snapshots ($L$) available is usually on the order of the number of sensors ($N$) in the array. In practice and in simulations it has been observed that the performance of the ABF depends on the ratio $N/L$ \cite{vtree08array}. 

Random Matrix Theory (RMT) offers an attractive framework to understand the
behaviors of SCMs.  RMT has results for the eigenstructure of
SCMs as the number of rows $N$ and columns $L$ of the data matrix go
to infinity while $N/L \rightarrow c$.
The resulting distributions are therefore characterized by the same
ratio that appears in ABF performance analysis.
Although RMT results are for the limiting case of infinitely large
matrices, they are frequently accurate for
modest data sizes.   This makes the RMT approach well
suited for analyzing the eigenvalue distribution of the SCM.

%

The Polynomial Method (PM) is an RMT technique for calculating the asymptotic eigenvalue distribution of a class of `algebraic' random matrices \cite{rao2008polynomial}. The Stieltjes transform of the \edf{} of algebraic random matrices satisfies a polynomial equation. The PM is based on a transform representation of a random matrix. Conceptually, this is similar to the Laplace transform used to represent scalar random variables by polynomial moment generating functions. Both techniques are based on a one-to-one correspondence between probability density functions (PDFs) and  polynomials. The Laplace transform represents the PDFs of a scalar random variable as a univariate polynomials, i.e., moment generating functions.  The PM requires several additional layers in its transform
representation whose details are well beyond the limited scope of the present paper. The central concept is that the PDF for the eigenvalues of a random matrix is represented by a bivariate polynomial. A set of deterministic and stochastic operations on random matrices are mapped to operations on the bivariate polynomials. The polynomial representations are thus manipulated in the manner corresponding to the desired operations on the random matrices. Finally, the polynomial representation is transformed back to the \edf{} of the desired output random matrix. The bivariate polynomial manipulations corresponding to common matrix operations can be quite complicated, but fortunately the toolbox RMTool is available to handle the symbolic algebra \cite{raormtool}.


This paper presents a method to predict an approximate \edf{} for the SCM of a \cin{} field. The technique presented here is similar in spirit to the results presented in \cite{ravi2012asympscm}, but it differs in two important ways. First, this paper focuses on \cin{} rather than the spherically isotropic noise in \cite{ravi2012asympscm}. Second, this paper exploits the PM and its RMTool toolbox rather than working directly with the Stieltjes transform as in \cite{ravi2012asympscm}.

The next section describes the method of applying the
PM to obtain an approximate EDF of the noise SCM. Sec.~\ref{sec:results} illustrates the application of this technique for a particular array size. Finally, Sec.~\ref{sec:conclusion} provides a short discussion of the results.

\section{Method}
\label{sec:method}

This section describes a technique to compute an approximate
\edf{} for the SCM of cylindrically isotropic noise observed by a
uniform linear array (ULA).  The technique exploits properties of
free multiplicative convolution \cite{rao2008polynomial} to
approximate the eigenvalue density of an $N\times N$ SCM by replacing
the \edf{} of the \ECM{} by an atomic density (PDF containing only Dirac delta functions) with fewer than $N$ atoms.  The
PM computes a numerical approximation to the SCM \edf{} using this
lower order atomic density.  

Let \Knn{} be the \ECM{} for the cylindrically isotropic noise
measured at the $N$-element ULA. The entries of this matrix are given
by \eqref{eqn:iso_noise_cov}. The eigenvalues of \Knn{} are $\eval_1 \geq
\eval_2 \geq \ldots \eval_N \geq 0$. The data matrix \datamtx{} is an
$N\times L$ matrix of complex phasors representing the $L$ temporally
independent but spatially correlated snapshots observed on the array
after demodulating to baseband.  These snapshots can be modeled as 
$\datamtx = \Knn^{1/2} \GNL$, where \GNL{} is an $N\times L$ matrix of
independent, identically distributed proper complex Gaussian
random variables with zero mean and unit variance.  This model
guarantees that the SCM converges to the desired \ECM{},
i.e. $E\{(1/L)\datamtx \datamtx\herm\} = \Knn$.  

The SCM is computed from the data matrix $\X$ as
\begin{equation}
\label{eq:scm}
\sampCov = (1/L)\X\X^H = (1/L)\Knn^{1/2}(\GNL\GNL^H)\Knn^{1/2}.
\end{equation}
The eigenvalues of \sampCov{} are $\seval_1 \geq \seval_2 \geq \ldots
\seval_N$. The SCM of \GNL{} is a Wishart matrix $\Wishart =
(1/L)\GNL\GNL^H$ where {$c=N/L$}.  Thus the SCM in \eqref{eq:scm} 
can be expressed as $\sampCov =  \Knn^{1/2} \Wishart
\Knn^{1/2n}$. This matrix has the same eigenvalues as the product
\Knn\Wishart{}. The Wishart matrix is an algebraic
matrix \cite[Remark 5.15]{rao2008polynomial} with an \edf{} given by
the \marpas{} (MP) density $\fMP{}(x)$ parameterized by $c$
\cite{marcenko1967distribution}.  If \Knn{} is an algebraic matrix,
then the product is also an algebraic matrix \cite[Theorem
5.19]{rao2008polynomial}.  Thus, if \Knn{} can be modeled as an
algebraic matrix, the PM provides a straightforward way to compute
the eigenvalue density for $\Knn\Wishart{}$, or equivalently, the
\edf{} for the SCM.  

The simplest way to create an algebraic density for \Knn{} is to
construct an atomic density with all $N$ eigenvalues of \Knn{}, each
with mass $1/N$.  All matrices with atomic eigenvalue densities fall
within the class of algebraic random matrices \cite[Example
3.6]{rao2008polynomial}.  The polynomial representation of the SCM
\sampCov{} ($\Lmz^{\sampCov}$) can be found directly from the
polynomials representing \Knn{} ($\Lmz^{\Knn}$) and the Wishart matrix
\Wishart{} ($\Lmz^{\Wishart}$) using the Multiply Wishart operation in
the PM \cite[Table 7]{rao2008polynomial}.  The dependence of the SCM
eigenvalues on the number of snapshots enters through the
parameterization of $\Lmz^{\Wishart}$.  An inverse operation is performed 
on  $\Lmz^{\sampCov}$ to extract the desired density $f^{\sampCov}(x)$ on 
the support region of interest \cite{raormtool}.  

The drawback of this approach is that the degree in $m$ of the
polynomial $\Lmz^{\sampCov}$ grows as $\mathcal{O}(N)$.
Moreover, the free multiplicative convolution (FMC) describing
\Knn{}\Wishart{} replaces each impulse in the atomic 
\edf{} of \Knn{} with some non-linearly convolved version of the MP 
density, i.e., $\fhat$, to produce the continuous eigenvalue density 
function for \sampCov.  The ensemble eigenvalues whose separation is
 much less than the support region the density $\fhat{}$ will be
 smeared together resulting into single continuous density. This 
suggests that the eigenvalue density of \sampCov{} can be modeled 
using many fewer than $N$ atoms for the density of \Knn{} by intelligently 
exploiting the smearing that results when multiplying \Knn{} by a Wishart
matrix.  As a result, the \edf{} $f^{\Knn}(x)$ generated by using all
$N$ atoms from \Knn{} can be replaced by a modified \edf{}
$\tilde{f}^{\Knn}(x)$ with many fewer atoms, resulting in a much lower
order polynomial $\Lmz^{\sampCov}$ substantially reducing computational time.

\begin{figure}[thb]
	\centerline{\includegraphics[width=1.0\linewidth]{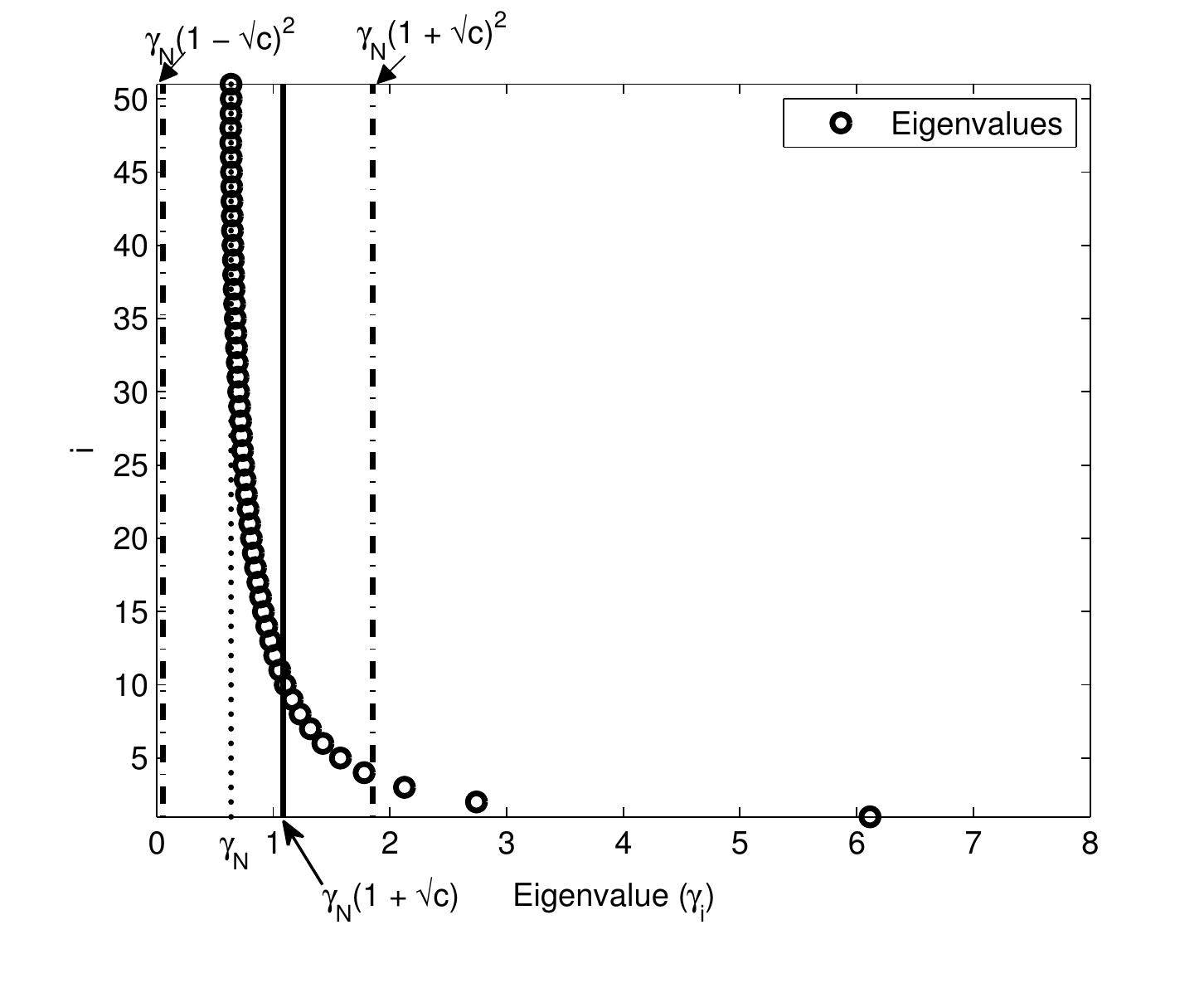}}
	\caption{Ensemble eigenvalues (circles) of \Knn{} computed for the case of $N = 51$ and $c = 0.25$. The support set of \marpas{} distribution is denoted by dashed lines and the threshold value is denoted by the solid line.}
	\label{fig:ensembEig}
\end{figure}

Designing the reduced order model relies on properties of the
covariance matrix $\Knn{}$ for cylindrically isotropic noise.   The
covariance matrix is a Hermitian Toeplitz matrix whose entries are
samples of $J_0(\alpha n)$. The eigenvalues of such a matrix are asymptotically equally distributed as the samples of the Fourier transform of the entries of the first row of \Knn{}. For \cin{}, the first row is $J_0(\alpha n)$ \cite{gray1972asymptotic,szego84toeplitz} and the Fourier
transform is equal to  $F(\omega) = 2/\sqrt{\alpha^2 - \omega^2}$ for $|\omega| < \alpha$.  The form of this Fourier transform
implies that most of eigenvalues will be very close to $2/\alpha$,thus very
 close together relative to the width of the resulting MP density. Only a small
 subset of eigenvalues will be sufficiently spaced to remain distinct 
after smearing by the MP PDF in the nonlinear FMC. Fig.~\ref{fig:ensembEig}
 shows the eigenvalues of \Knn{} for N = 51, where the 
eigenvalues are plotted on the horizontal axis against their index on the
 vertical. This behavior is very similar to what is known as a spiked covariance
 model in RMT.
 
The SCM eigenvalue behavior for a spiked covariance model is
described in \cite{paul2007asymptotics}. This model assumes that the data matrix 
consists of a low rank perturbation in a unit power white noise background.  In
the event that the white noise background is not unit power, it is
straightforward to scale the problem by the eigenvalue $\eval_N$
representing the background power.  Assuming $\eval_N=1$, the
$N_{low}$ ensemble eigenvalues between
\Threshold{} and 1 will produce $N_{low}$ SCM eigenvalues
$\seval_{N-N_{low}+1},\ldots,\seval_{N_{low}}$ distributed
nearly indistinguishably than if there had been a single atom at $1$ with
mass $N_{low}/N$ \cite{paul2007asymptotics}.  
This suggests that all ensemble eigenvalues $\eval_i
\leq \Threshold$ can be collapsed into a single atom at $\eval_N = 1$
with mass $N_{low}/N$ without significant impact on the SCM eigenvalue
 distribution.  This atom will be replaced by the non-linearly convolved version of MP density $\fhat{}$, in the EDF of $\sampCov$.
 The eigenvalues with $\eval_i  > \Threshold$ will behave as distinct atoms in principle. However, many of these atoms are also very closely spaced relative to the width of support width of $\fhat$ and will also
be smeared together nearly indistinguishably in the density for \sampCov.  Consequently, these atoms are also collapsed into a single atom at $\eval_{mid} =
 (\Threshold + \WishartUL)/2$. Finally the eigenvalues above
 \WishartUL{} maintain their identity as distinct atoms. 

%
 
To define the model precisely, let $\Gamma_{dist} = \{\eval_i |\eval_i > \WishartUL\} $ 
be a set of atoms expected to remain distinct even after FMC. The number of eigenvalues
in different ranges are given by $N_{mid}  = |\{\eval_i|\Threshold < \eval_i< \WishartUL \}|$ 
and $N_{low} = |\{ \eval_i | 1 \leq \eval_i \leq \Threshold\}|$ where $|\cdot|$ indicates 
the cardinality of the set. Then the modified \edf{} for \Knn{} is 
\begin{align}
\tilde{f}^{\Knn{}}(x) & = \frac{1}{N}\sum\limits_{\eval_i\in\Gamma_{dist}} \delta(x - \eval_i) + \notag \\ 
  & \frac{N_{mid}}{N} \delta(x - \eval_{mid}) + \frac{N_{low}}{N} \delta(x - \eval_N)
\label{eqn:eval_model}
\end{align}
The SCM eigenvalue density $f^{\sampCov}(x)$ can be computed using \eqref{eqn:eval_model} and
the multiplication by Wishart properly as described earlier. 

This approach results in a much lower order polynomial
$\Lmz^{\sampCov}$ to represent \sampCov{}. For the example in
Fig.~\ref{fig:ensembEig}, this approach reduces the atomic distribution
from $N=51$ to a mere $5$ atoms. The computation required in solving for the roots of the polynomial $\Lmz^{\sampCov}$ is of the order  $\mathcal{O}(N^3)$ \cite{edelman1995polynomial}. Hence the lowered polynomial degree results in substantial savings in computational requirement. 



%

\section{Simulation Results}
\label{sec:results}
\begin{figure}[thb]
    \centerline{\includegraphics[width=1.0\linewidth]{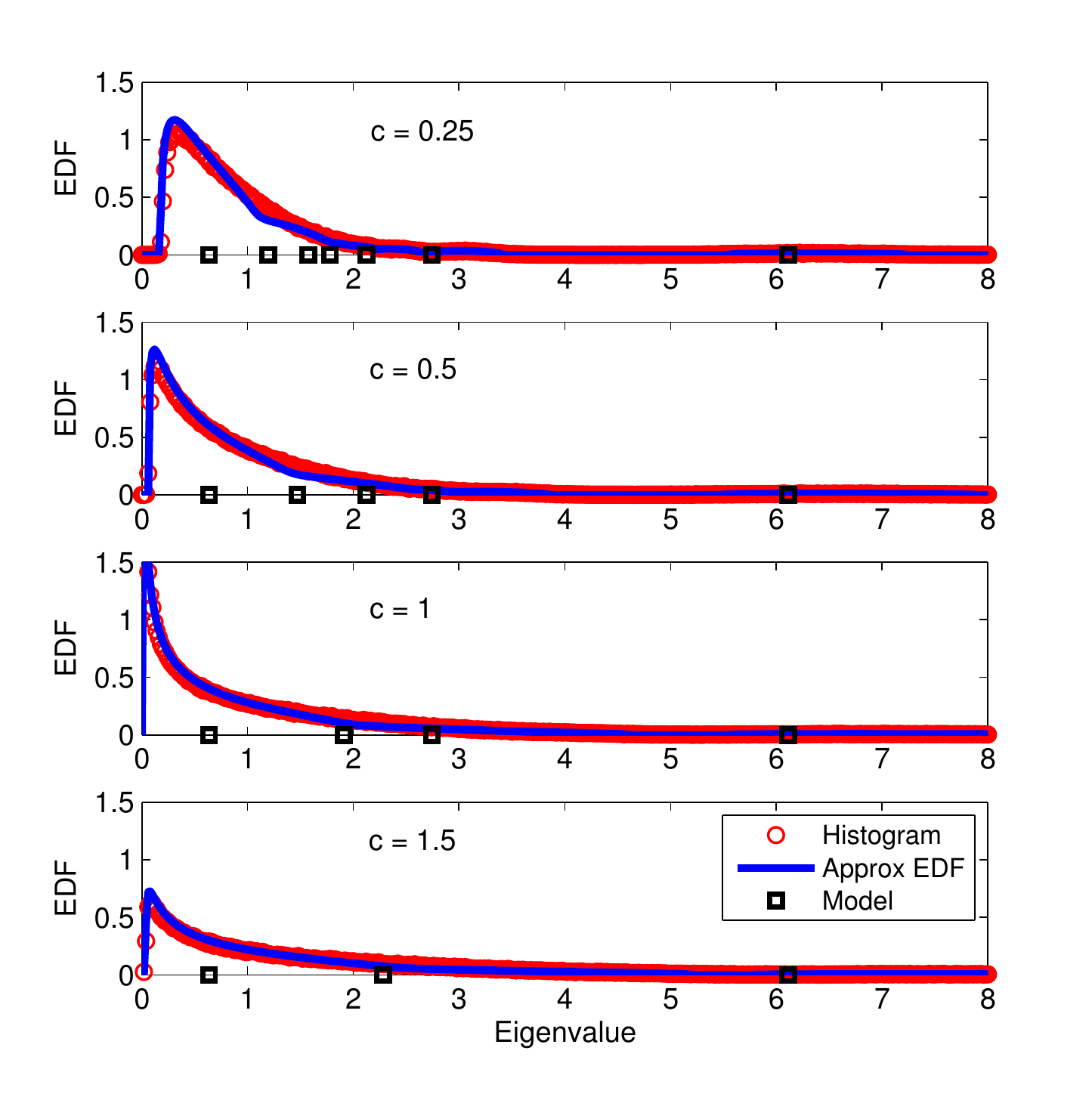}}
	\caption{Comparison of approximate \edf{} (solid blue) with histogram of eigenvalues (red circles) observed from 5000 Monte Carlo simulations of N = 51 element horizontal uniform linear array. The location of atoms for the EDF in \eqref{eqn:eval_model} are shown as black squares in each panel. The four panels shows different scenario from snapshot rich case at the top ($L = 204, c = 0.25$) to snapshot deficient case at the bottom ($L = 34, c = 1.5$).}
		\label{fig:histKI}
\end{figure}

This section compares the SCM eigenvalue density predicted by the model described in Sec.~\ref{sec:method} with histograms obtained through Monte Carlo simulations of \cin{} measured at a horizontal ULA with $N = 51$ sensors at $\lambda/2$ spacing. The approximate \edf s  obtained for the SCM are compared with simulation results for different numbers of snapshots to verify the accuracy of the technique. Fig.~\ref{fig:histKI} compares the \edf{} predicted by the method in Sec.~\ref{sec:results} with a histogram obtained from 5000 Monte Carlo simulations. 

Fig.~\ref{fig:ensembEig} shows the ensemble eigenvalues (circles) for \Knn. Note that most of the eigenvalues are clustered around the smallest eigenvalue $\eval_N = 2/\pi = 0.6366$ and a few eigenvalues are distinctly larger than the rest. As mentioned in the Sec.~\ref{sec:method},  \Knn{} can be viewed as a spiked covariance matrix, most of whose eigenvalues are $\eval_N = 0.6366$. Note that because the smallest eigenvalue is not one as in the canonical spiked covariance model, the threshold and support regions for the model must all be scaled by $\eval_N$ when determining the atomic distribution. Thus, the two dashed lines in Fig.~\ref{fig:histKI} indicate the upper and the lower limit of the \marpas{} density scaled by $\eval_N$, and the solid line indicates the scaled threshold value. Note that there are at least three dominant ensemble eigenvalues, one well separated at around $\eval_1 = 6.11$ and two slightly separated around $\eval_2 = 2.74$ and $\eval_3 = 2.12$. 

Fig.~\ref{fig:histKI} shows a comparison of the approximate \edf{} for the SCM and the histograms. The blue line indicates the approximate \edf{} computed using the PM, while the red circles indicate the histogram from the simulation. The four panels correspond to $c = \{0.25, 0.5, 1, 1.5\}$ from top to bottom, respectively. The choice of values for $c$ covers a range of sensor to snapshot ratios that describe many practical scenarios.

In all four cases, there is close agreement between the \edf{} and the simulation histograms, suggesting that this method for approximating the \edf{} of the \cin{} SCM is accurate.  Note that for the cases with $c>1$, \sampCov{} is singular thus, the eigenvalue density also includes an impulse of area $(1-1/c)$ at $x=0$ that is not shown on these
figures. 

\section{Discussion and Conclusion}
\label{sec:conclusion}
The simulation results in Fig.~\ref{fig:histKI} confirm that the approximate \edf{} computed from the method in Sec.~\ref{sec:method} gives a good approximation of the histogram of the eigenvalues obtained from the simulation.  

In practice this algorithm is limited by the symbolic computation of the roots of \Lmz{} for the Stieltjes transform $m(z)$ for the SCM \cite{raormtool}. As noted in Sec.~\ref{sec:method}, the degree of the polynomial \Lmz{} in $m$ grows with the number of atoms in the model density function \eqref{eqn:eval_model}. From \eqref{eqn:eval_model} it is evident that the eigenvalues below \WishartUL{} always contribute two atoms. But the eigenvalues above \WishartUL{} contribute as distinct atoms. The number of eigenvalues modeled as distinct atoms depends on the choice of $c$ and $N$. As the order of $\Lmz^{\sampCov}$ grows, the number of roots to be solved for also grows.

This model can be combined with signal models to produce more accurate estimates of ABF performance for bearing estimation in shallow water where the background noise is often cylindrically isotropic.  Additionally, as discussed in \cite{ravi2012asympscm}, understanding the nature of the isotropic noise model will make it clear when noise eigenvalues will appear as distinct in $\Gamma_{dist}$, and should prevent misinterpretation of these noise eigenvalues as false targets.

In conclusion, the proposed method approximates the EDF for the SCM of cylindrically isotropic noise using the PM to realize a substantial computational savings. The method exploits properties of FMC to model the SCM EDF with a greatly reduced polynomial order. This results in a lower order polynomial \Lmz{} hence less computation is required to solve for its roots. The \edf{} obtained from this method gives a good approximation of the histogram of eigenvalues obtained from simulation.



\bibliographystyle{IEEEbib}
\bibliography{refs}

\begin{thebibliography}{10}

\bibitem{vtree08array}
H.~L. Van~Trees,
\newblock {\em Optimum Array Processing},
\newblock Wiley-Interscience, 2002.

\bibitem{cox1973spatial}
H.~Cox,
\newblock ``Spatial correlation in arbitrary noise fields with application to
  ambient sea noise,''
\newblock {\em J. Acoust. Soc. Am.}, vol. 54, pp. 1289--1301, 1973.

\bibitem{kuperman80spatial}
W.~A. Kuperman and F.~Ingenito,
\newblock ``Spatial correlation of surface generated noise in a stratified
  ocean,''
\newblock {\em J. Acoust. Soc. Am.}, vol. 67, no. 6, pp. 1988--1996, 1980.

\bibitem{rao2008polynomial}
N.~R. Rao and A.~Edelman,
\newblock ``The polynomial method for random matrices,''
\newblock {\em Foundations of Computational Mathematics}, vol. 8, no. 6, pp.
  649--702, 2008.

\bibitem{raormtool}
N.~R. Rao,
\newblock ``{RMT}ool: A random matrix and free probability calculator in
  {MATLAB},'' http://www.eecs.umich.edu/~rajnrao/rmtool/.

\bibitem{ravi2012asympscm}
R.~Menon, P.~Gerstoft, and W.~Hodgkiss,
\newblock ``Asymptotic eigenvalue density of noise covariance matrices,''
\newblock {\em IEEE Trans.Sig. Proc}, 2012.

\bibitem{marcenko1967distribution}
VA~Mar{\v{c}}enko and L.A. Pastur,
\newblock ``Distribution of eigenvalues for some sets of random matrices,''
\newblock {\em Mathematics of the USSR-Sbornik}, vol. 1, pp. 457, 1967.

\bibitem{gray1972asymptotic}
R.~Gray,
\newblock ``On the asymptotic eigenvalue distribution of {T}oeplitz matrices,''
\newblock {\em IEEE Trans. Info. Th.}, vol. 18, no. 6, pp. 725--730, 1972.

\bibitem{szego84toeplitz}
U.~Grenander and G.~Szeg{\H{o}},
\newblock {\em Toeplitz forms and their applications},
\newblock Chelsea. Pub. Co., 1984.

\bibitem{paul2007asymptotics}
D.~Paul,
\newblock ``Asymptotics of sample eigenstructure for a large dimensional spiked
  covariance model,''
\newblock {\em Statistica Sinica}, vol. 17, no. 4, pp. 1617--1642, 2007.

\bibitem{edelman1995polynomial}
A.~Edelman and H.~Murakami,
\newblock ``Polynomial roots from companion matrix eigenvalues,''
\newblock {\em Mathematics of Computation}, vol. 64, no. 210, pp. 763--776,
  1995.

\end{thebibliography}

\end{document}